\documentclass[12pt]{article}
\usepackage[dvips]{graphicx,psfrag}
\usepackage{amssymb}
\input epsf
\newcommand{\be}{\begin{equation}}
\newcommand{\ee}{\end{equation}}
\newcommand{\bea}{\begin{eqnarray}}
\newcommand{\eea}{\end{eqnarray}}
\newcommand{\lb}{\label}
\begin{document}
\begin{titlepage}
\begin{center}
{\large\bf  PRIMORDIAL BLACK HOLES FROM INFLATIONARY MODELS
             WITH AND WITHOUT BROKEN SCALE INVARIANCE}
\vskip 1cm
{\bf Torsten Bringmann}
\vskip 0.4cm
 Fakult\"at f\"ur Physik, Universit\"at Freiburg,\\
 Hermann-Herder-Str.~3, 79104 Freiburg, Germany.
\vskip 0.7cm
{\bf Claus Kiefer}
\vskip 0.4cm
 Institut f\"ur Theoretische Physik, Universit\"at zu K\"oln,\\
 Z\"ulpicher Str.~77, 50937 K\"oln, Germany.
\vskip 0.7cm
{\bf David Polarski}
\vskip 0.4cm
 Laboratoire de Physique Math\'ematique et Th\'eorique,
 UMR 5825 CNRS,\\
 Universit\'e de Montpellier II, 34095 Montpellier, France.\\
\vskip 0.3cm
 Laboratoire de Math\'ematiques et Physique Th\'eorique,
 UMR 6083 CNRS,\\
  Universit\'e de Tours, Parc de Grandmont, 37200 Tours, France.\\

\end{center}
\date{\today}
\vskip 2cm
\begin{center}
{\bf Abstract}
\end{center}

\begin{quote}
We review the formalism of primordial black holes (PBHs) production and
show that the mass variance at horizon crossing has been systematically 
overestimated in previous studies. We derive the correct expression.
The difference is maximal at the earliest formation times and still very 
significant for PBH masses $\sim 10^{15}$g, an accurate estimate 
requiring numerical calculations. In particular, this would lead to weaker 
constraints on the spectral index $n$.
We then derive constraints on inflationary models from the fact that
primordial black holes must not overclose the Universe.
This is done both for the scale-free case of the power spectrum
studied earlier and for the case where a step in the mass variance
is superimposed.
In the former case we find various constraints on $n$,
depending on the parameters. In the latter
case these limits can be much more strengthened, so that
one could find from an observational limit on $n$ a constraint
on the allowed height of the step.
\end{quote}

\end{titlepage}
 
	\section{Introduction}
Cosmology has now entered an exciting stage 
where observations of ever increasing
accuracy can probe the remote past of our Universe.
The present paradigm makes use
of an inflationary stage of expansion in the very early Universe
(see e.g. \cite{Li90,LL,pad}). This solves
some of the big problems of the old big bang cosmology
in an elegant way,
in particular the problems of horizon and flatness. But most importantly,
and there lies its predictive power, it gives a
possible solution to the crucial problem of where the primordial
fluctuations leading to the observed large-scale structure (LSS) come from.
In fact, they have their origin in the ubiquitous vacuum fluctuations.
The seed of the LSS has been
observed in the form of tiny fluctuations imprinted on the
cosmological microwave background (CMB) at the time of decoupling.
Each inflationary model makes precise predictions 
about the spectrum of its primordial fluctuations and this is how these
models can be constrained by observations.
 
It was realized already some time ago that primordial fluctuations can
lead to the formation of primordial black holes (PBHs)
(see e.g. \cite{Carr85} for a review).
For this, one needs a spectrum of primordial fluctuations whose statistics and
amplitude are known. Inflationary models produce a
spectrum of fluctuations also on scales relevant for the formation of PBHs.
The statistics for PBH formation is usually taken 
as Gaussian for simplicity, which
fits most of the inflationary models.
One therefore possesses an additional tool
to constrain inflationary models. The a priori interesting thing about PBHs is
that they probe scales which are many orders of magnitude smaller than scales
probed by large-scale structure (LSS) surveys and 
CMB angular anisotropy observations. In this sense it is
analogous, even if less spectacular, to the primordial gravitational wave
background of inflationary origin.

It was already shown that PBH formation can put constraints on the spectral
index of the primordial density perturbations, $n\le (1.23-1.25)$,
see e.g. \cite{CGL,KL96,GL97,BK01}. Basically, these constraints
come either from the evaporation of black holes
(leading to a somewhat stronger limit) or from the fact that
PBHs must not overclose the present Universe, i.e., from
the requirement $\Omega_{PBH,0}<1$. On mass scales $M\lesssim
10^{35}\ \textrm{g}$,
these constraints on $n$ are even stronger than those coming from COBE.
There are also constraints from the need not to overproduce
PBHs during the preheating phase following inflation \cite{BT,GM00}.
Such constraints assume of course that the primordial spectrum 
is scale-free, 
an assumption that we shall relax here. 
Hence there are two attractive aspects in the study of PBHs using inflationary
perturbations: it expands the range of scales that can be probed, and it can
test possible features in the spectrum on these scales
not probed by CMB and LSS observations.
 
A characteristic scale can be generated, for example, in two-fields inflation
such as double inflation (see e.g. \cite{PS92,P94}) or in single-field 
inflation
with broken scale invariance (BSI) (see e.g. \cite{St92,LPS}).
Such a model was used recently in order to drastically cut the power on
small scales in an attempt to explain the dearth of dwarf galaxies
\cite{KL00}.
We want here to extend this study to the formation of PBHs.
We are, therefore, interested in features on much smaller scales
than those which we have considered previously, 
an exciting possibility which we seriously consider
as there is no reason why the spectrum should
be scale-free on all those scales relevant for PBH formation
(though the actual occurrence of the scale-free case
 cannot be excluded). 

Our paper is organized as follows. In Sect.~II we start by reviewing the
basic formalism for PBH formation. We then point out in detail 
how expressions for the mass variance used earlier have to be corrected
in order to obtain accurate numbers. 
We then consider several concrete models, beginning with the
usual scale-free case, then extending our study to the presence
of a step in the mass variance. In Sect.~III we then obtain
constraints on the spectral index from the requirement
that PBHs must not overclose the Universe. For the scale-free case
we get various constraints on $n$, depending on the parameters of 
interest, while for the case of a scale-free spectrum with a 
superimposed step the constraint on $n$ depends on the height of the
step. Therefore, one can obtain constraints on this height from the
existing limits on $n$. Finally, Sect.~IV contains our conclusions.

\section{PBH formation}
\subsection{Basic formalism}
 
A spectrum of primordial density
fluctuations can lead to the production of primordial black holes. 
Let us assume for simplicity that a PBH is formed when the density contrast
averaged over a volume
of the (linear) size of the Hubble radius
 satisfies $\delta_{min} \leq \delta \leq
\delta_{max}$,
and the PBH mass, $M_{PBH}$, is of the order of the
``horizon mass'' $M_H$, the mass contained inside the Hubble volume.
Usually one takes $\delta_{min}=\frac{1}{3},~\delta_{max}=1$ coming from
semianalytic considerations \cite{Carr85}.
 Recent numerical calculations indicate,
however, that PBHs can be formed over a much wider range of masses
at a given formation time and that $\delta_{min}\approx 0.7$ \cite{NJ,GL99}.
This is why we leave the limits open at this stage. For the purpose of
studying the changes introduced by a characteristic scale
 in the primordial spectrum,
it will be sufficient in the following to make the simplifying
assumption that PBHs are formed with the mass $M_H$.
We shall, however, also comment below on the changes introduced
by the assumption that the PBH mass is only a certain fraction of $M_H$.
 
Each physical scale $R(t)$ is defined by some wavenumber $k$ and evolves with
time according to
$R(t)= {a(t)}/{k}$.
For a given physical scale $R(t)$, the ``horizon'' crossing time $t_k$ -- here
we do not
mean ``horizon'' crossing during inflation, but after inflation -- is the time
when this scale reenters
the Hubble radius, which will inevitably happen after inflation for scales that
are greater than the
Hubble radius at the end of inflation. It is at this time $t_k$
where the above condition for the density contrast holds. It will
lead to the formation of a PBH with mass
 $M_{PBH}$, being approximately equal to
$M_H(t_k)$. Clearly, there is a one-to-one correspondence between $R(t_k),
M_H(t_k)$, and $k$.
Of course we can also take this correspondence at any other initial time $t_i$
and relate the physical quantities at both times $t_i$ and $t_k$.
 
Generally, if the primordial fluctuations obey Gaussian
statistics, the probability density $p_R(\delta)$, where $\delta$ is the density
contrast averaged over
a sphere of radius $R$, is given by
\be
p_R(\delta) = \frac{1}{\sqrt{2\pi}~\sigma (R)}~ e^{-\frac{\delta^2}{2
\sigma^2(R)}}\ .
\ee
Here, the dispersion (mass variance) $\sigma^2(R)\equiv \Bigl \langle \Bigl (
\frac{\delta M}{M}
\Bigr )_R^2 \Bigr \rangle$ is computed using a top-hat window function
\cite{LL,pad},
\be
\sigma^2(R) = \frac{1}{2\pi^2}\int_0^{\infty}\textrm{d}k\;
 k^2 ~W^2_{TH}(kR) ~P(k)\lb{sigW}~,
\lb{int}
\ee
where $P(k)$ is the power spectrum. It is defined by
$P(k)\equiv\langle|\delta_k|^2\rangle$ in the case of finite volume.
For inflation, one has to take instead the
infinite-volume expression 
\be
\langle \delta_{\mathbf k} \delta^*_{{\mathbf k}'}
 \rangle \equiv P(k)~ \delta ({\mathbf k}-{\mathbf k}')\ ,
\ee
where we assume isotropy of the ensemble. 
From a fundamental point of view, the averages $\langle\ldots\rangle$
refer to quantum expectation values; however, an effective 
quantum-to-classical transition is achieved during 
inflation \cite{PS96,KPS}. 
 In the case of PBHs produced by 
inflationary perturbations, this quantum-to-classical transition 
is guaranteed for the masses of interest to us (see \cite{P01} and the remarks 
at the end of section 2.2).

The expression
$W_{TH}(kR)$ stands for the Fourier transform of the top-hat window
function divided by the probed volume $V_W=\frac{4}{3}\pi R^3$,
\be
W_{TH}(kR)=\frac{3}{(kR)^3}\bigl (\sin kR-kR\cos kR\bigr )\ .
\end{equation}

Therefore the probability $\beta(M_H)$ that a region
 of comoving size $R=k^{-1}$ has an averaged 
density contrast at horizon crossing $t_k$ in the range 
$\delta_{min}\leq\delta\leq\delta_{max}$, is given by
\be
\label{beta}
  \beta(M_H)=\frac{1}{\sqrt{2\pi}\,\sigma_H(t_k)}\,
           \int_{\delta_{min}}^{\delta_{max}}\,
           e^{-\frac{\delta^2}{2 \sigma_H^2(t_k)}}\,\textrm{d}\delta
          \approx\frac{\sigma_H(t_k)}{\sqrt{2\pi}\,\delta_{min}}
           e^{-\frac{\delta_{min}^2}{2 \sigma_H^2(t_k)}}\ ,
\ee
where $\sigma_H^2(t_k):=\sigma^2(R)\big|_{t_k}$, and the last approximation
is valid for $\delta_{min}\gg\sigma_H(t_k)$, and
$(\delta_{max}-\delta_{min})\gg\sigma_H(t_k)$.

Important conclusions can be drawn from (\ref{beta}). 
Let us consider first the value of $\beta(M_H)$ 
today. We then have $\sigma_H^2(t_0)\simeq 10^{-8}$. 
So clearly the probability of forming a black hole 
today is extraordinarily small. This probability
 can be larger in the primordial universe if the 
power is increased when we go backwards in time,
 but usually the number will remain 
very small, $\beta(M_H)\ll 1$, at all times.
This is due to the amplitude of ${\delta^2_{min}}/{\sigma^2_H(t_k)}$,
on which $\beta(M_H)$ depends sensitively, see (\ref{beta}).

The expression (\ref{beta}) for $\beta(M_H)$ is usually interpreted as
giving the probability that a PBH will be 
formed with a mass $M_{PBH}\geq M_H(t_k)$, i.e. greater or
equal to the mass contained inside the Hubble volume when this scale
reenters the Hubble radius. Strictly speaking,
 though, this is \emph{not} true, since (\ref{beta}) 
does not take into account those regions that are underdense on a scale $M_H$,
 but nevertheless overdense on some larger scale. 
In the Press-Schechter formalism (see e.g. \cite{LL,pad})
 this seems to be taken care of
in some models by multiplying (\ref{beta}) with a factor $2$. 
Fortunately, as emphasized in \cite{GL97},
 in most cases $\beta(M_H)$ is a very rapidly falling function 
of mass, so this effect can be neglected. 
In this case, $\beta(M_H)$ \emph{does} give the probability
 for PBH formation and thus also 
(at time $t_k$) the mass fraction of regions that will evolve
 into PBHs of mass greater or equal to 
$M_H$. This will also be the case for our models.

\subsection{Improved calculation of the mass variance}
           \label{secalpha}

As is clear from the previous subsection, one is
interested in computing $\sigma^2(R)$ at horizon crossing using the
power spectrum of the primordial fluctuations $P(k,t)$ of interest. 
Let us consider the simplest case which is usually considered when
the power spectrum is scale-free and of the form $P(k,t)=A(t)~k^n$.
Then, the following equation is usually used to relate
$\sigma_H(t_k)$ to the present value $\sigma_H(t_0)$, see e.g.
\cite{GL97}, 
\be
\sigma^2_H(t_k) = \sigma^2_H(t_0)~\Biggl \lbrack \frac{M_H(t_0)}
{M_H(t_{eq})} \Biggr \rbrack^{\frac{n-1}{3}}
                  \Biggl \lbrack  \frac{M_H(t_{eq}}{M_H(t_k)}  \Biggr
\rbrack^{\frac{n-1}{2}}\ .
\lb{sig}
\ee
The expression (\ref{sig}) rests on the assumption
that $\sigma_H(t_k)$ is well approximated by the 
integral (\ref{int}), without window function
 but with a cut-off at the horizon scale. 
Consequently, the presence of the window function
 would be equivalent to the introduction of an 
effective cut-off at the horizon scale, leading to 
(recalling that $A(t_k)\propto k^{-4}$)
\be
\sigma^2_H(t_k)\propto k^{n-1} \ .\lb{sc}
\ee
However, for a scale-free primordial spectrum with $n\geq 1$,
one recognizes that (\ref{int}) is 
ultraviolet divergent! Hence it is clear that the integral is dominated
 by the contribution from small scales. 
Actually, even for $0.8<n<1$, the small scales will still
 contribute significantly to the integral and even dominate it
 for $n\approx 1$.
   
On the other hand, the quantities $k^{3}~\Phi^2(k,t)$,
 or $\delta^2_H(k,t)$, defined as 
\be
\delta^2_H(k,t)\equiv \frac{(aH)^4}{k^4}\frac{k^3}{2\pi^2}~P(k,t) \equiv
          \frac{2}{9 \pi^2}~k^3~\Phi^2(k,t)
\ee 
are time-independent on ``superhorizon'' scales
 (scales bigger than the Hubble radius, $k<aH$)
in the cases that we consider here, being
equal in very good approximation to their value
 at the horizon crossing time $t_k$. On these scales,
these quantities behave as in (\ref{sc}).
In addition, the constant of proportionality
 is different in the matter and radiation 
dominated stages. With $t_{k_r}$ and $t_{k_m}$ denoting times 
 in the radiation and matter dominated stage, respectively, we get
(cf. e.g. \cite{PS92,LL})
\bea
k_r^{3}~\Phi^2(k_r,t_{k_r}) &=&
 \left ( \frac{10}{9}\right )^2 ~k_m^{3}~\Phi^2(k_m,t_{k_m}) 
              \left ( \frac{k_r}{k_m} \right )^{n-1}~,\nonumber \\
\delta^2_H(k_r,t_{k_r}) &=& \left ( \frac{10}{9}\right )^2
        ~\delta^2_H(k_m,t_{k_m})
              \left (\frac{k_r}{k_m} \right )^{n-1}~.\lb{sca}
\eea
However, inspection of (\ref{beta}) shows that
 it is the quantity $\sigma_H(t_k)$ which is needed. 

Actually (and fortunately), there is a natural upper cut-off
 in $k$-space for the power spectrum, 
namely $k_e$, corresponding to the Hubble radius at the end of inflation $t_e$.
The lower limit 
can be taken zero if we assume that the number
 of e-folds during inflation amply solves 
the cosmological particle horizon problem.
We thus have to consider only wavelengths bigger than $k_e^{-1}$.
The relation between $\sigma_H(t_k)$ and $k^{\frac{3}{2}}~\Phi(k,t_k)$, 
or $\delta_H(t_k)$, {\it depends on the time $t_k$ and is model-dependent}. 
We have in general (with $k=(aH)|_{t_k}$) 
\be
\sigma^2_H(t_k) 
   \equiv \alpha^2(k)~\delta^2_H(k,t_k)~,\lb{alpha1}
\ee
where
\be
\alpha^2(k)= \int_0^{\frac{k_e}{k}} x^{n+2}~
 T^2(kx,t_k)~W^2_{TH}(x)~\textrm{d}x~.\lb{alpha}
\ee
The transfer function $T(k,t)$ is defined through
\be\lb{T}
P(k,t)= \frac{P(0,t)}{P(0,t_i)}~P(k,t_i)~T^2(k,t)\; , \quad T(k\to 0,t)\to 1~.
\ee
Here, $t_i$ is some initial time when all scales
 are outside the Hubble radius, $k\ll aH$, and we can take $t_{i}=t_e$.

The problem in evaluating $\alpha(k)$ comes from the evolution 
of the perturbations for scales $k'$ inside the 
Hubble radius, $k=(aH)|_{t_k}\leq k' \leq k_e$, 
or equivalently $1\leq x \leq \frac{k_e}{k}$.
 This small scale evolution is encoded in the 
transfer function $T(k,t)$ which is defined through relation (\ref{T}).
For scales outside the horizon, the transfer function equals one.
We emphasize that
in (\ref{alpha}), the transfer function must be taken
 {\it at the time $t_k$} of interest, not today. 
Clearly, an accurate value for (\ref{alpha})
 can be obtained only numerically and with an explicit knowledge
of $T(k,t)$, which is of course model dependent \cite{LL}.

Let us now compare our results with what is usually done using (\ref{sig}).
The quantity $\sigma_H(t_0)$ is certainly finite,
 even for $n\geq 1$, because one must use 
in (\ref{int}), according to (\ref{alpha}), the actual power spectrum whose deformation 
on small scales is encoded in the present-day transfer function $T(k,t_0)$,
\be
\sigma^2_H(t_0) = \delta^2_H(t_0)~\int_0^\infty
 x^{n+2}~W^2_{TH}(x)~T^2(k_0 x,t_0)~\textrm{d}x~, 
\lb{sigT}
\ee
where $k_0\equiv (aH)_{t_0}$, and $\delta_H(t_0)\equiv\delta_H(k_0,t_0)$. 
The transfer function today, $T(k,t_0)$, effectively cuts the power on small scales and makes 
(\ref{sigT}) finite.
In addition, {\it today} $T(k,t_0)$ becomes already negligible for scales $k\ll k_e$; this is why 
the upper limit of integration in (\ref{sigT}) can be replaced by infinity.
 
The ratio between $\sigma_H(t_0)$ and $\delta_H(t_0)$ often used
(see e.g. \cite{GL97,KL96}),
\be
\sigma_H(t_0)\approx 5~\delta_H(t_0),\lb{sig0}
\ee
when simply substituted into (\ref{sig}), would yield the result
\be
\sigma^2_H(t_k) \approx 25~\delta^2_H(t_0)~\Biggl \lbrack \frac{M_H(t_0)}
{M_H(t_{eq})} \Biggr \rbrack^{\frac{n-1}{3}}
                  \Biggl \lbrack  \frac{M_H(t_{eq})}{M_H(t_k)}  \Biggr
\rbrack^{\frac{n-1}{2}}\ .
\lb{sigin}
\ee
Eq. (\ref{sigin}) is a reasonable first guess, but too inaccurate
 if one is willing 
to make precise predictions. 
According to (\ref{sca}) and (\ref{alpha1}),
 one must instead use for $t_k\ll t_{eq}$ the expression
\bea
\sigma^2_H(t_k) &=& \frac{100}{81} ~\alpha^2(k)~\delta^2_H(t_0)~\Biggl \lbrack \frac{M_H(t_0)}
{M_H(t_{eq})} \Biggr \rbrack^{\frac{n-1}{3}}
                  \Biggl \lbrack  \frac{M_H(t_{eq})}{M_H(t_k)}  \Biggr
\rbrack^{\frac{n-1}{2}}\ ,\nonumber \\
&=& \frac{200}{9^3 \pi^2}~\alpha^2(k)~k_0^{3}~\Phi^2(k_0,t_0) ~\Biggl \lbrack \frac{M_H(t_0)}
{M_H(t_{eq})} \Biggr \rbrack^{\frac{n-1}{3}}
                  \Biggl \lbrack  \frac{M_H(t_{eq})}{M_H(t_k)}  \Biggr
\rbrack^{\frac{n-1}{2}}~.\lb{sigb1}
\eea
Eqs. (\ref{sigb1}), (\ref{alpha1}), and (\ref{alpha}) are the main 
results of this section.
Eq. (\ref{sigin}) corresponds, in our notation,
 to $\alpha(k)=\alpha(k_0)=5\times 9/10=4.5$.
It is not correct as $\alpha(k)$ is both model and scale dependent,
since the transfer 
function depends on the cosmological parameters
and the time $t_k$. 
Even for a given cosmological model for which (\ref{sig0}) holds, it would certainly 
be incorrect at much earlier times $t_k$, according to (\ref{alpha}). 
This can result in a large 
discrepancy in the computation of $\beta(M_H)$. It is recognized that the problem is much 
deeper than just a clever choice of the window function. 
Actually, the window function corresponding to the numerical results obtained 
in \cite{NJ} is the top-hat window function. If one uses $\delta_{min}\approx 0.7$ 
as found by these authors, one must also use the top-hat window function 
in order to be consistent with their numerical results. 

It is tacitly assumed in (\ref{sigb1}),
 (as in (\ref{sig})), that the PBHs form 
before the time of equality, $t_k\ll t_{eq}$, and that the universe is first 
radiation-dominated and instantaneously becomes matter-dominated at $t_{eq}$ 
(with $\Lambda=0$). A more complicated evolution of the scale factor $a(t)$ 
will give a different expression.
The observational input is the present density contrast 
$\delta_H(t_0)$ on the Hubble-radius scale
which is found using the CMB anisotropy data.

What are the scales for which the discrepancy between (\ref{sigb1}) 
and (\ref{sigin}) is most important? To get an idea, we consider 
the earliest formation times $t_k\sim t_{k_e}$, right after inflation.
This gives
\be
\alpha^2(k_e) \approx \int_0^1 x^{n+2}~W^2_{TH}(x)~\textrm{d}x \approx 0.21~, 
\lb{inte}
\ee
with 
\be
0.202 \leq \alpha^2(k_e) \leq 0.218~~~~~~~~~~~{\rm for}~~~~~~~ 1.3\geq  n \geq 1~.  
\ee
Clearly there is a big difference for the earliest formation time $t_k\sim t_{k_e}$.
A first rough estimate of $\alpha(k)$ corresponding to $M_H\approx 10^{15}$g 
gives $(\frac{10}{9})^2~\alpha^2(M_H\approx 10^{15}{\rm g})
\lesssim 2$, substantially smaller than $25$. This estimate follows
by taking into account that for modes outside the horizon one has
$|\delta_k|\propto a^2$, while for modes within the horizon 
$|\delta_k|$ is oscillating.
. 

Interestingly, as seen from (\ref{beta}),
 the change implied by a smaller $\sigma_H(t_{k})$ at earlier 
times, {\it amplifies} the change induced by a larger $\delta_{min}$ --
one is led to a smaller $\beta(M)$. 
Therefore, constraints on $n_{max}$ from PBH formation
both on mass scales $M<10^{15}$g (these concern PBHs that have already evaporated by today) 
as on scales $M>10^{15}$g (see below) will 
be {\it less constraining} if one uses (\ref{sigb1})
 instead of (\ref{sigin}).
We note finally that an effective quantum-to-classical transition 
takes place already for PBHs with masses $M\ll 10^{15}$g
, but not for PBHs with $M\sim 1$g 
corresponding to $k\sim k_e$ \cite{P01}
(with the exception of PBHs possibly produced during preheating, see e.g. \cite{BT,GM00}). 
At these formation times, 
the growing and decaying modes of the fluctuations
 that reenter the Hubble radius 
are still of the same order (no squeezing). Therefore, 
the classical approach must be amended. Anyway, a correct expression of 
the power spectrum must then take the decaying mode into account \cite{LL00}. 
  
\subsection{Inflationary spectra with a scale}

It is also our intention to investigate what happens
 when the assumption of a scale-free spectrum, 
tacitly or implicitly done in the literature, is dropped.
We therefore consider spectra with a characteristic scale
 for which (\ref{sca}) and therefore
(\ref{sigb1}) do not apply anymore. For each specific spectrum we have to compute $\sigma_H(t_k)$
numerically, but it is possible to give a rough estimate of what will happen 
if we consider some fiducial cases. 
Consider for example the case of a spectrum with a pure step in $\sigma_H(t_k)$ at $k=k_s$, 
with $t_{k_s}<t_{eq}$, where the ratio of $\sigma_H(t_k)$ on large scales to
that on small scales is given by $p$. In this case, we have
\be
\label{sig2}
  \sigma_H(t_{k})= \frac{10}{9}~\alpha(k) ~\delta_H(t_0)
     \Biggl\lbrack\frac{M_H(t_0)}{M_H(t_{eq})}\Biggr \rbrack^{\frac{n-1}{6}}
     \Biggl\lbrack\frac{M_H(t_{eq}}{M_H(t_k)}\Biggr\rbrack^{\frac{n-1}{4}}\times
     \left\{ \begin{array}{ll} 
                  1      & \textrm{for $k<k_s$} \\ 
                  p^{-1} & \textrm{for $k\geq k_s$} 
     \end{array} \right.\ .
\ee

We emphasize that, as explained in Sect.~2, this does not correspond 
to a pure step in the primordial power spectrum $\delta_H(t_k)$.
But with the approximation of a constant $\alpha$ for all PBH masses greater than $10^{15}$g, 
this is certainly the simplest extension to (\ref{sigb1}) that 
has a characteristic scale. 
Here we shall restrict ourselves to this case.
In a future paper we shall then consider models with a characteristic scale which are 
motived by an underlying Lagrangian, or even features in the inflaton potential, 
for example a particular primordial fluctuations spectrum which derives from a jump 
in the inflaton potential derivative. The resulting fluctuations spectrum for such a potential 
is of a universal form and an exact analytical expression has been derived by 
Starobinsky \cite{St92}.
This expression depends (in addition to the overall normalization) 
on two parameters $p$\ and $k_s$, which have a similar meaning to those in the 
toy model considered here.

In our future work we will also take into account the effect of a 
cosmological constant corresponding to 
$\Omega_{\Lambda}\approx 0.7$ at present, since this seems to be 
strongly supported by recent observations and makes our current 
comprehension of the universe converge into a coherent picture.

\section{Observational constraints}

There are various limits on the initial mass fraction
 $\beta(M)$ of PBHs \cite{GL97}, 
most of them being related to the effects of Hawking radiation. 
Here, we restrict ourselves to the gravitational constraint
\begin{equation}
  \label{gc}
  \Omega_{PBH,0} = 2 ~\Omega_{PBH,eq}~\Omega_{m,0} < 1\ ,
\end{equation}
which states that the present PBH mass density must not exceed the present density of 
the universe. 
In (\ref{gc}), $\Omega_{m,0}$ refers to {\it all} dustlike matter today
including a possible contribution from PBHs.
In order to relate (\ref{gc}) to the initial PBH abundance, one needs in 
principle to know the entire history of the universe right from the 
time when a mass scale $M$ crosses into the horizon. 
Then one can compare the evolution of the PBH density $\rho_{PBH}$ 
(which goes like $\rho_{PBH}\propto a^{-3}$) to the total density $\rho_{tot}$. 
In a radiation dominated universe $\rho_{tot}\propto a^{-4}$, so a small 
initial fractional PBH density can grow very large until today. 
%
%We assume the ``standard'' evolution, i.e. a radiation dominated universe 
%until the recent matter-dominated era. 
%
We take again the evolution of the scale factor corresponding to (\ref{sigb1}).
We follow
steps similar to \cite{GL97}, but take into account all $g$-factors,
i.e., use the constancy of the entropy $S=g_{*S}(T)a^3T^3$. This leads to 
the limit
\begin{equation}
  \label{gc2}
  \beta(M_H) < 1.57 \times 10^{-17}\left(\frac{M_H}{10^{15}\textrm{g}}\right)^
\frac{1}{2}~h^2,
\end{equation}
where we have used $t_{eq}=4.36\times10^{10}\textrm{ s}\,(\Omega_{m,0}h^2)^{-2}
\times (\frac{T_{\gamma}}{2.75})^6$ \cite{KT90}. Strictly speaking, the behaviour 
of the scale factor is slightly more sophisticated, but we leave that aside here. 
Anyway, as we will show in Fig.1, a small change in (\ref{gc2}) does not matter.
Finally, since black holes of initial mass $M_{PBH}<5\times10^{14}$ g will have 
evaporated by the present day, the tightest constraint is obtained for $M_H\sim10^{15}$ g.

\subsection{Scale-free spectrum}

\begin{figure}[t]
  \begin{center} 
     \psfrag{n}[][][0.8]{$n$}
     \psfrag{lg beta}[][][0.8]{$\lg \beta$}
     \psfrag{a1}[][][0.6]{$\alpha=4.5$}
     \psfrag{a2}[][][0.6]{$\alpha=3$}
     \psfrag{a3}[][][0.6]{$\alpha=1.5$}
  \includegraphics[width=0.7\textwidth]{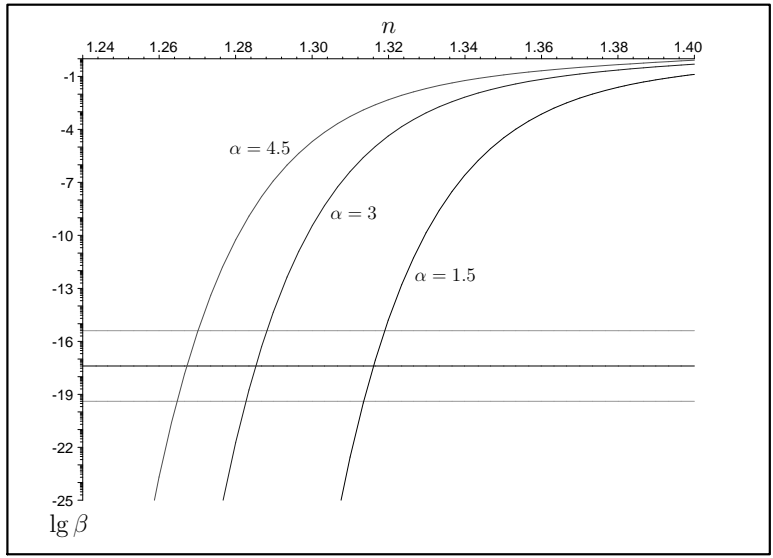} 
  \end{center}
  \caption[]{The quantity $\beta\equiv\beta(M=10^{15}\mbox{ g})$
 is shown as a function of the spectral index $n$, for several values
 of $\alpha\equiv\alpha(M= 10^{15}\mbox{ g})$.
 The straight lines represent the gravitational constraint (\ref{gc2})
 and, for the purpose of illustration, the same constraint weakened and
 strengthened by a factor of $100$, respectively. The resulting maximally
 allowed value for $n$ clearly depends only weakly on the precise value of
 the prefactor of the gravitational constraint (\ref{gc2}), whereas it
 depends rather sensitively on the value of $\alpha$. The value usually
 used in the literature, $\alpha\approx 4.5$, overestimates the initial PBH
 abundance significantly and thus leads to much stronger constraints on
 $n$ (i.e. $n<1.27$) than would be expected for the more realistic choice
 of $\alpha\approx 1.5$ (see section \ref{secalpha}), which results in
 $n\lesssim 1.32$.}
  \label{betan}
\end{figure}

In order to compare the constraints on the spectral index
 with earlier results we first consider a scale-free spectrum, 
$n={\rm const}$, on all scales.  
 We further make the approximation $\alpha(k)\approx \alpha(k_0)
\approx 4.5$ in order to compare our results with previous work.
%
%As we have seen in Sect.~II, this is in general not quite correct.
%
As we have seen in Sect.~II, this is not correct.
The exact numbers can only be calculated numerically.
This is beyond the scope of our paper, but we shall get an idea about
the changes by displaying the dependence
of the obtained constraints on the used value for $\alpha(k)$, see
Fig.~\ref{betan}.

The data obtained from COBE can be used \cite{GL97} to normalize the spectrum (\ref{sig}) to 
\begin{equation}
  \label{s0}
  \sigma_H(M_{H,0}\sim 10^{56}\textrm{ g})=9.5\times10^{-5}\ .
\end{equation}
The above numbers are understood as to include the factor ${10}/{9}$.
 In this way, the comparison with 
earlier results is straightforward.
 With the following setting of the parameters,
\begin{equation}
  \label{par}
  h:=0.5,\qquad\delta_{min}:=\frac{1}{3},
\end{equation}
and by using equations (\ref{beta}) and (\ref{gc2}), evaluated at $M_H=10^{15}$ g, one gets for the 
spectrum (\ref{sig}) the following upper bound on the spectral index:
\begin{equation}
  \label{n1}
  n<1.27.
\end{equation}
This result differs from $n<1.31$ \cite{GL97} and is much closer to the constraints they found due to the 
evaporation of PBHs, $n<1.24$. This justifies our approach of
only looking at the gravitational constraint - 
particularly since the details of black hole evaporation are still quite speculative. 
As mentioned above, the real constraints (i.e., taking into account
the full expression (\ref{alpha}) for $\alpha$) are expected to be
somewhat weaker, cf. Fig.~\ref{betan}.

We also had a look at the dependence of the result on the particular choice of the 
parameters (\ref{par}). 
What we found was a weak dependence on $h$ and $M_H$ (i.e. the minimal mass to which the 
gravitational constraint (\ref{gc2}) applies): varying $h$ from $0.1$ to $1$ as well as varying $M_H$ 
from $10^{14}$ g to $10^{16}$ g changes the constraint on $n$ by only about $0.01$.

The motivation for varying $M_H$ is twofold:
 first, it amounts to the fact that the details of black 
hole evaporation are not yet fully understood,
 so the minimal mass of PBHs that have not evaporated 
by the present day is afflicted with some (small) uncertainty; second, if the mass of the PBH formed 
is not equal to the horizon mass $M_H$,
 but somewhat smaller, i. e. $M_{PBH}=\epsilon M_H, \epsilon<1$, 
the minimal horizon mass to which the gravitational constraint (\ref{gc2}) applies, is changed to 
$M_H=\epsilon^{-1}\cdot 10^{15}$ g. Semianalytical considerations, for example, require 
$\epsilon=\gamma^\frac{3}{2}$ \cite{Carr85}. 
In the radiation dominated case ($\gamma=\frac{1}{3}$) this would weaken the constraint (\ref{n1}) 
by only about $0.005$. But even $\epsilon=0.001$ weakens it by only slightly more than $0.02$!

\begin{figure}[t]
  \begin{center} 
     \psfrag{n_{max}}[][][0.8]{$n_{max}$}
     \psfrag{\\delta_{min}}[][][0.8]{$\delta_{min}$}
     \includegraphics[width=0.7\textwidth]{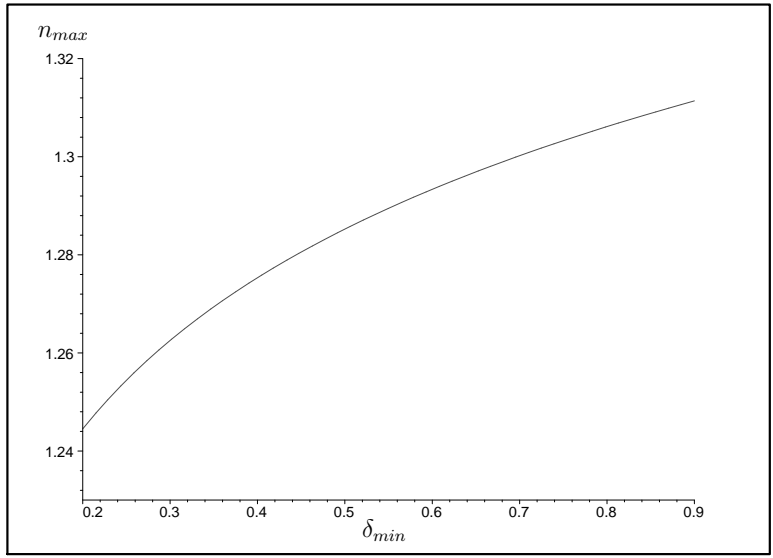} 
  \end{center}
  \caption[]{Dependence of the constraint on the spectral index,
 $n_{max}$, on $\delta_{min}$ for the case of a scale-free spectrum, with 
$h=0.5, M_H=10^{15}$ g. The constraint on $n$ is clearly weakened
 for larger values of 
$\delta_{min}$.} 
  \label{dmin}
\end{figure}

On the other hand, we have found, as expected, the dependence
 on the value of $\delta_{min}$ to be 
rather strong: for $\delta_{min}=0.2$ one gets $n<1.24$ and for $\delta_{min}=0.7$ one gets $n<1.30$. 
The latter point is rather important, since $\delta_{min}=\frac{1}{3}$ (as assumed above) relies 
on semianalytical arguments, whereas numerical analysis seems to support $\delta_{min}\approx0.7$
\cite{NJ}.
 Fig.~\ref{dmin} shows the dependence of $n_{max}$ on $\delta_{min}$. 

\subsection{Scale-free spectrum with a superimposed step}
 
Next we consider the case of a simple extension to the scale-free spectrum, 
i.e. a power-law spectrum with a step as described by (\ref{sig2}). 
Again, we use the approximation $\alpha(k)\approx\alpha(k_0)\approx 4.5$
in order to compare our results with the previous subsection.
We restrict ourselves to the case $p\leq 1$, i.e. more power on smaller scales. 
The opposite case leads to interpretational problems for the quantity $\beta(M)$: 
For $p>1$ it is no longer a monotonically falling function of mass and therefore 
cannot be interpreted anymore as the fractional mass density of PBHs of mass 
larger than $M$. For such a case, the Press-Schechter formalism
would have to be modified.
 But $p<1$ is the physically interesting case for us,
since it would produce {\em more} instead of less PBHs on
smaller scales.

\begin{figure}[b]
  \begin{center}
     \psfrag{lg \\beta(M)}[][][0.7]{$\lg \beta(M)$}
     \psfrag{lg M [g]}[][][0.7]{$\lg M$ [g]}
     \includegraphics[width=0.7\textwidth]{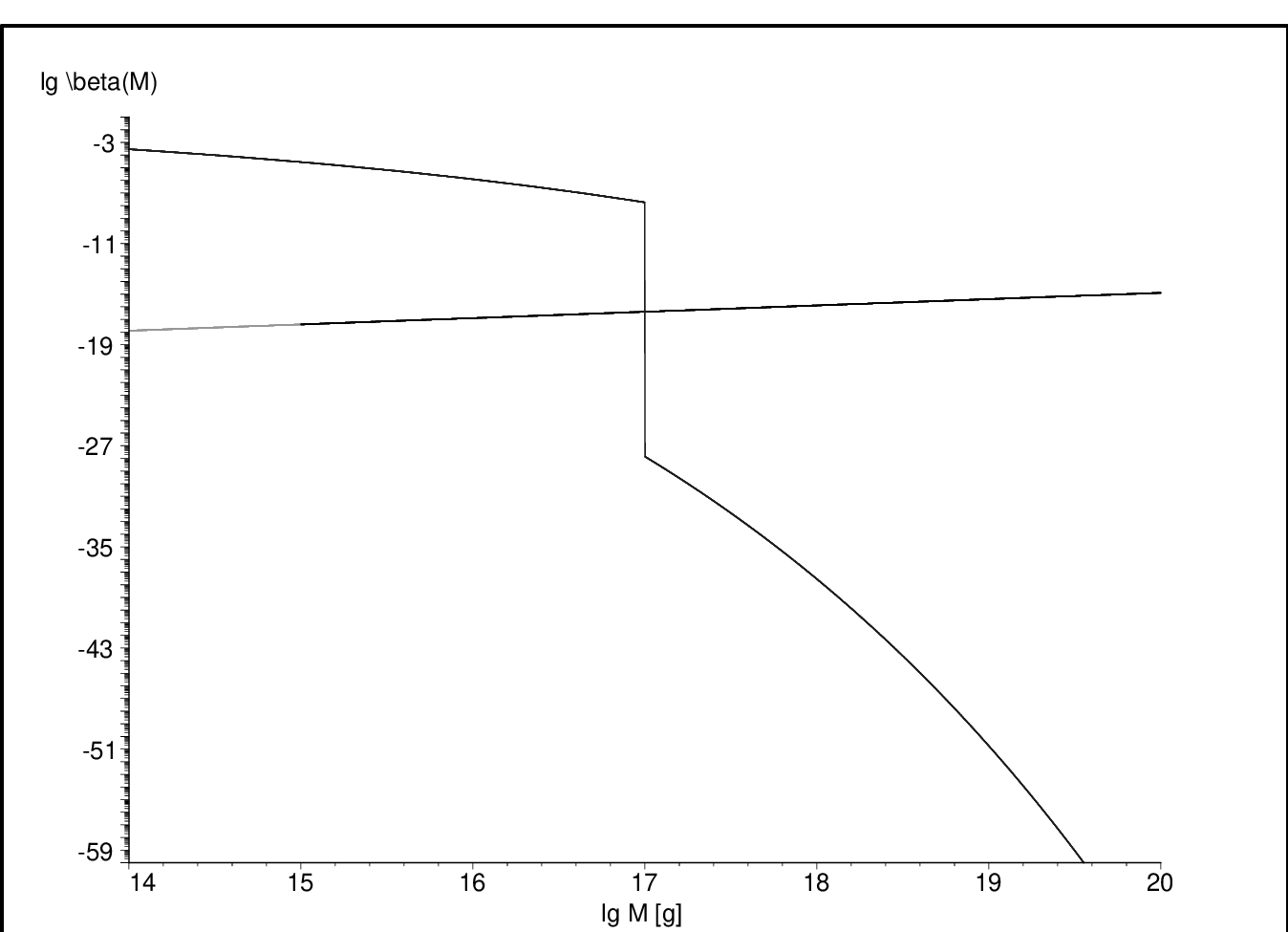} 
  \end{center}
  \caption[]{This figure shows $\beta(M)$ for a step spectrum as
 in (\ref{sig2}), 
with (arbitrarily chosen) $n=1.27$, $M_H(t_{k_s})=10^{17}$ g and $p=0.5$.
 The straight line 
is the gravitational constraint (\ref{gc2}), which applies only for
 $M\gtrsim 10^{15}$ g. The figure illustrates that the
 gravitational constraint is to be evaluated at $10^{15}$ g
 and that the result does not depend on the choice of $k_s$
 (as long as $M_H(t_{k_s})>10^{15}$ g).} 
  \label{betafig}
\end{figure}

In contrast to the scale-free case we now have three parameters - $n$, $p$ and 
$k_s$ (or $M_H(t_{k_s})$, respectively) - to be determined. One would therefore 
generally expect that the constraint (\ref{gc2}) on $\beta(M)$ gives a plane 
in this parameter space. Since this constraint is a rising and $\beta(M)$ a 
falling function of mass, the strongest constraint
 will as before come from  
$M\sim 10^{15}$ g. But as long as $M_H(t_{k_s})>10^{15}$ g, 
$\beta(M\sim10^{15}\textrm{ g})$ does not depend on the choice of $k_s$, 
so we get a functional dependence between $n_{max}$ and $p$ only. 
In order to illustrate this point, Fig. \ref{betafig} shows $\beta(M)$ 
together with the gravitational constraint (\ref{gc2}).

\begin{figure}[t]
  \begin{center} 
     \psfrag{lg p}[][][0.8]{$\lg p$}
     \psfrag{n_{max}}[][][0.8]{$n_{max}$}
     \includegraphics[width=0.7\textwidth]{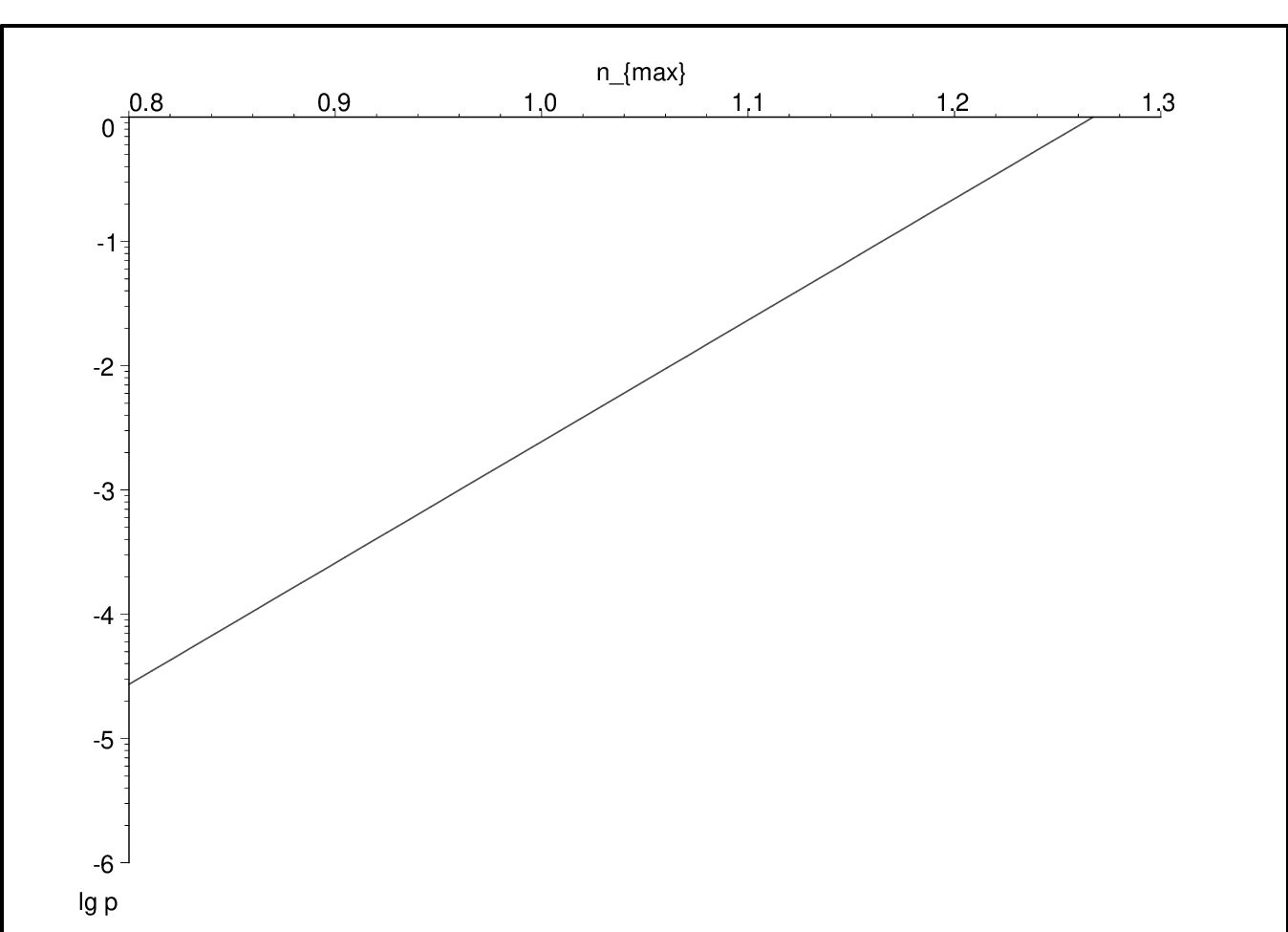} 
  \end{center}
  \caption[]{Dependence of $n_{max}$ on $p$ for the step spectrum 
(\ref{sig2}). 
For $p=1$ the result for the scale-free case is recovered. For $p<1$ the 
constraint on $n$ is considerably strengthened.} 
  \label{pn}
\end{figure}

Applying the same 
calculations as described in the previous paragraph, one gets $n_{max}(p)$ 
as depicted in Fig.~\ref{pn}.
For $p=1$ the result of the previous subsection is recovered (i.e. $n_{max}=1.27$), 
but for lower values of $p$ the constraint on $n$ may be considerably strengthened. 
For example, $p=0.01$ requires $n<1.06$ ! Of course this result can also be used the 
other way around: if $n>1.06$ were found by some experimental measure, the 
gravitational constraint on the PBH abundance would require $p>0.01$. 

The analysis sketched above does not tell us anything about the scale $k_s$ at 
which the step occurs. It would nevertheless be interesting to 
learn something about $k_s$. This could arise,
 for example, from observations of MACHOs, 
hinting at the presence of a feature in $\beta(M)$ in the mass range $M > 10^{15}$g. 

\section{Conclusions}
We have considered two aspects of the formation of PBHs.
 On the one hand, we have 
derived a more precise expression for the PBH abundance. 
It corrects the usual formula used, and could in principle substantially  
improve the accuracy of the obtained results. 
The amount of correction implied would need numerical calculation
and will be presented elsewhere.
Our results concerning this part of the present work are summarized in 
(\ref{sigb1}), (\ref{alpha1}), and (\ref{alpha}). 
On the other hand, we have considered simple models where 
the spectrum is not scale-free and have compared it to the scale-free case. 
In the latter case we have found a somewhat stronger constraint
than in \cite{GL97} and have discussed
the dependence of the result on the value of $\delta_{min}$.
In the former case,  
we have found that the presence of a step in the mass variance
can drastically strengthen the constraints on $n$, giving in turn
a tool to constrain the height of the step. 
This study can be extended to models with an underlying 
miscroscopic Lagrangian or phenomenologically motivated 
features in the potential. We plan to address these issues in
a future publication. 

\section*{Acknowledgements}
We are grateful to Andrew Liddle and Karsten Jedamzik for enlightening 
discussions.


\begin{thebibliography}{99}
\bibitem{Li90} A. D. Linde, {\em Particle physics and inflationary cosmology} 
(Harwood, New York, 1990).
\bibitem{LL} A.R. Liddle and D.H. Lyth, {\em Cosmological inflation
and large-scale structure} (Cambridge University Press, Cambridge, 2000).
\bibitem{pad} T. Padmanabhan, {\em Structure formation in the
 Universe} (Cambridge University Press, Cambridge, 1993).
\bibitem{Carr85} B.J. Carr, in {\em Observational and theoretical
aspects of relativistic astrophysics and cosmology}, edited by
J.L. Sanz and L.J. Goicoechea (World Scientific, Singapore, 1985).
\bibitem{CGL} B.J. Carr, J.H. Gilbert, and J.E. Lidsey,
Phys. Rev. D {\bf 50}, 4853 (1994).
\bibitem{KL96} H.I. Kim, C.H. Lee,
   and J.H. MacGibbon, Phys. Rev. D {\bf 59}, 063004 (1999).
\bibitem{GL97} A.M. Green and A.R. Liddle, Phys. Rev. D {\bf 56},
 6166 (1997).
\bibitem{BK01} E.V. Bugaev and K.V. Konishchev,
       {\tt astro-ph/0103265}. 
\bibitem{BT} B.A. Bassett and S. Tsujikawa, Phys. Rev. D {\bf 63}, 123503
             (2001).
\bibitem{GM00} A. M. Green and K.A. Malik,
         Phys. Rev. D {\bf 64}, 021301 (2001); 
 F. Finelli, S. Khlebnikov, Phys. Lett. B {\bf 504}, 309 (2001).
\bibitem{PS92} D. Polarski and A.A. Starobinsky,
  Nucl. Phys. B {\bf 385}, 623 (1992).
\bibitem{P94} D. Polarski, Phys. Rev. D {\bf 49}, 6319 (1994).
\bibitem{St92} A.A. Starobinsky, JETP Lett. {\bf 55}, 489 (1992).
\bibitem{LPS} J. Lesgourgues, D. Polarski, and A.A. Starobinsky,
 Mont. Not. Roy. Astron. Soc. {\bf 297}, 769 (1998).
\bibitem{KL00} M. Kamionkowski and A.R. Liddle, 
 Phys. Rev. Lett. {\bf 84}, 4525 (2000).
\bibitem{NJ} J.C. Niemeyer and K. Jedamzik, Phys. Rev. Lett. {\bf 80},
 5481 (1998); Phys. Rev. D {\bf 59}, 124013 (1999).
\bibitem{GL99} A.M. Green and A.R. Liddle, Phys. Rev. D {\bf 60},
               063509 (1999).
\bibitem{PS96} D. Polarski and A.A. Starobinsky,
 Class. Quantum Grav. {\bf 13}, 377 (1996).
\bibitem{KPS} C. Kiefer, D. Polarski, and A.A. Starobinsky,
 Int. J. Mod. Phys. D {\bf 7}, 455 (1998).
\bibitem{P01} D. Polarski, Int. J. Mod. Phys. D, to appear (2001); 
e-print Archive astro-ph/0109388
\bibitem{LL00} S.M. Leach and A.R. Liddle,
 Phys. Rev. D{\bf 63}, 043508 (2001).
\bibitem{KT90} E. Kolb and M. Turner, {\em The early Universe}
 (Addison-Wesley, Redwood City, 1990).


\end{thebibliography}
\end{document}